\documentclass[aps,prd,showkeys,nofootinbib,floatfix,amsmath,amssymb,showpacs,superscriptaddress,notitlepage]{revtex4-1}

\usepackage{rotating}
\usepackage{graphicx}
\usepackage{dcolumn}
\usepackage{bm}
\usepackage{color}
\usepackage{multirow}
\usepackage{mathtools}
\usepackage[pdftex,bookmarks=true]{hyperref}
\hypersetup{
   colorlinks,
   citecolor=black,
   filecolor=black,
   linkcolor=black,
   urlcolor=black
}

\begin{document}

\title{\texorpdfstring{$\Xi_c \gamma \rightarrow\Xi^\prime_c$}{Xc gamma -\> X'c} transition in lattice QCD}

\author{H. Bahtiyar}
\affiliation{Department of Physics, Yildiz Technical University, Davutpasa Campus Esenler 34210 Istanbul
Turkey}
\affiliation{Department of Physics, Mimar Sinan Fine Arts University, Bomonti 34380 Istanbul Turkey}
\author{K. U. Can\footnote{Current address: RIKEN Nishina Center, RIKEN, Saitama 351-0198, Japan}}
\affiliation{Department of Physics, H-27, Tokyo Institute of Technology, Meguro, Tokyo 152-8551 Japan}
\author{G. Erkol}
\affiliation{Department of Natural and Mathematical Sciences, Faculty of Engineering, Ozyegin University, Nisantepe Mah. Orman Sok. No:34-36, Alemdag 34794 Cekmekoy, Istanbul Turkey}
\author{M. Oka}%
\affiliation{Department of Physics, H-27, Tokyo Institute of Technology, Meguro, Tokyo 152-8551 Japan}
\affiliation{Advanced Science Research Center, Japan Atomic Energy Agency, Tokai, Ibaraki, 319-1195 Japan}
\author{T. T. Takahashi}%
\affiliation{Gunma National College of Technology, Maebashi, Gunma 371-8530 Japan}

\date{\today}

\begin{abstract}
We evaluate the electromagnetic $\Xi_c \gamma \rightarrow\Xi_c^\prime$ transition on 2+1 flavor lattices corresponding to a pion mass of $\sim 156$~MeV. We extract the magnetic Sachs and Pauli form factors which give the $\Xi_c$-$\Xi_c^\prime$ transition magnetic moment and the decay widths of $\Xi_c^\prime$ baryons. We did not find a signal for the magnetic form factor of the neutral transition $\Xi_c^0 \gamma \rightarrow\Xi_c^{\prime 0}$, which is suppressed by the U-spin flavor symmetry. As a byproduct, we extract the magnetic form factors and the magnetic moments of $\Xi_c$ and $\Xi_c^\prime$ baryons, which give an insight to the dynamics of $u/d$, $s$ and $c$ quarks having masses at different scales.

\end{abstract}
\maketitle

\section{Introduction}
Recent experimental observations of all the ground-state heavy baryons as predicted by the quark model~\cite{Olive:2016xmw} makes it timely to study the structure and decays of these hadrons with theoretical methods. Among heavy baryons, $\Xi_c$ and $\Xi_c^\prime$ are particularly interesting as the three quarks they are composed of ($u$, $s$ and $c$) have different flavors and masses at quite different scales. Therefore, these two baryons provide a good laboratory to study the heavy-quark dynamics.

The neutral $\Xi_c^0$(c[sd]) and the positive state $\Xi_c^+$ (c[su]) have the quantum numbers $J^P =\frac{1}{2}^+$ and an anti-symmetric flavor wavefunction under interchange of light quarks. In group theoretical formalism they are members of the anti 4-plet ($\overline{4}$) of the SU(4) structure. $\Xi_c$ baryon was first observed in hyperon-beam experiment at CERN \cite{Biagi:1983en} and later confirmed by Fermilab \cite{Coteus:1986ar} and CLEO Collaboration~\cite{Alam:1989qt}. The average mass reported by PDG is $m_{\Xi_c^0} = 2470.99^{+0.30}_{-0.50}$~MeV \cite{Olive:2016xmw}.

The two other baryons with the same quark content and quantum numbers, $J^P =\frac{1}{2}^+$, are $\Xi'^0_c$ (c\{sd\}) and $\Xi'^+_c$ (c\{su\}), which are located on the second layer of the sextet SU(4) multiplet. These two baryons have symmetric flavor wavefunctions under interchange of light quarks. They were first observed by CLEO Collaboration~\cite{Jessop:1998wt} and confirmed recently by BABAR~\cite{Aubert:2006rv} and BELLE experiments~\cite{Yelton:2016fqw}. The average mass reported by PDG is $m_{\Xi'^0_c} = 2577.9 \pm 2.9$ MeV \cite{Olive:2016xmw}. 

The mass difference between $\Xi_c'$ and $\Xi_c$ first reported by CLEO as $\Delta M^+  = (107.8 \pm 1.7 \pm 2.5)$ MeV/$c^2$ and  $\Delta M^0  = (107.0 \pm 1.4 \pm 2.5)$ MeV/$c^2$~\cite{Jessop:1998wt} is too small for any strong decay occur. Therefore the electromagnetic $\Xi_c^\prime \rightarrow \Xi_c \gamma $ is the dominant decay mode. Studying this electromagnetic transition between different multiplets of SU(4) may shed light on the QCD mechanism governing the charmed baryons.

The experimental facilities such as LHCb, PANDA, Belle II, BESIII and J-PARC are expected to give more detailed information about spectroscopy, decays and structure of the charmed baryons. Concurrently, recent lattice-QCD studies provide a precise determination of their spectroscopy. The ground state charmed baryons have been studied both in quenched \cite{Lewis:2001iz,Mathur:2002ce} and full QCD \cite{Namekawa:2013vu,Alexandrou:2012xk,Alexandrou:2014sha,Briceno:2012wt}. 

In this work, we evaluate the $\Xi_c \gamma \rightarrow\Xi^\prime_c$ transition in 2+1-flavor lattice QCD. As a by-product we compute the electromagnetic form factors of $\Xi_c$ and $\Xi_c^\prime$ baryons. We make our simulations with near physical light-quark masses which give a pion mass of $\sim 156$~MeV. Using an appropriate ratio of two- and three-point correlation functions, we extract the electric and magnetic form factors which give the decay width of $\Xi_c^\prime$ and the magnetic moments of $\Xi_c$ and $\Xi_c^\prime$. A particular emphasis is made on the $\Xi^0_c \gamma \rightarrow\Xi^{0\prime}_c$ transition which is suppressed by the U-spin flavor symmetry. We find no signal on the lattice for this neutral transition. The electromagnetic decays of charmed baryons have been previously studied in heavy-hadron chiral perturbation theory \cite{Jiang:2015xqa,Banuls:1999br,Cheng:1992xi,PhysRevD.49.5857,*PhysRevD.55.5851.2}, quark models \cite{Dey:1994qi,Ivanov:1999bk,Ivanov:1996fj}, QCD sum rules \cite{Aliev:2016xvq}, bag model \cite{Bernotas:2013eia}, heavy-quark symmetry \cite{Tawfiq:1999cf} and lattice QCD~\cite{Bahtiyar:2015sga}. Preliminary results of this work are given in Ref.~\cite{Erkol:2017ewk}.

\section{Theoretical Formalism}
\subsection{Lattice Formulation}
We start with the definition of the electromagnetic vector current. We use the following electromagnetic current to study the electromagnetic and transition form factors of $\Xi_c$ and $\Xi_c'$ baryons: 
\begin{equation}
J_\mu =  \frac{2}{3} \overline{c}(x) \gamma_\mu c(x) - \frac{1}{3} \overline{s}(x) \gamma_\mu s(x) + c_\ell \overline{\ell}(x) \gamma_\mu \ell(x),	
\end{equation}
where $\ell$ denotes the flavor of the light quark (u and d) and $c_\ell$ is its charge ($2/3$ or $-1/3$). Using this definition of the vector current, we couple the current to each valence quark in the baryon and compute the electromagnetic transition form factors, which is described by the matrix element
\begin{equation}
 \langle {\cal B}^\prime(p',s')|J_\mu(x_1)|{\cal B}(p,s) \rangle  = \overline u(p',s') \Big[ \gamma _\mu F_1(q^2) - \frac {\sigma_{\mu \nu} q_\nu}{m_B+m_{B'}} F_2(q^2) \Big] u(p,s).
\end{equation}
Here $F_1(q^2),F_2(q^2)$ are the Dirac and Pauli form factors respectively. $u(p',s')$ and $u(p,s)$ are the Dirac spinor of the outgoing and incoming baryons with masses $m_{B'}$ and $m_B$, and $q_\mu = p'_\mu - p_\mu$ is the transferred four momentum with the momentum of the incoming (outgoing) baryon $p$ ($p'$).

The electric and magnetic Sachs form factors are defined in terms of Dirac and Pauli form factors as follows:
\begin{equation}\label{elS}
G_E(q^2) = F_1(q^2) - \frac{q^2}{(m_B+m_{B'})^2} F_2(q^2),
\end{equation}
\begin{equation}\label{magS}
G_M(q^2) = F_1(q^2) + F_2(q^2).
\end{equation}
We extract the form factors considering the following two-point correlation functions
\begin{equation}
\langle F^{{\cal B}\,{\cal B}}(t;p;\Gamma_4)\rangle=\sum\limits\limits_{x}^{ }\,e^{i\,p\,x} \Gamma^{\beta\,\alpha}_4\,\langle\Omega|T(\chi_{\cal B}^{\beta}(x) \overline \chi_{\cal B}^{\alpha}(0))|\Omega \rangle,
\label{eq:twopt}
\end{equation}
and the following three-point correlation functions 
\begin{equation}
\langle F^{{\cal B}^{\prime} J_\mu {\cal B}}(t_2,t_1;p',p;\Gamma) \rangle=-i\,\sum\limits\limits_{x_1,x_2}^{ }\,e^{-i\,p\,x_2}\,e^{i\,q\,x_1} \Gamma^{\beta\,\alpha}\,\langle\Omega|T(\chi^{\beta}_{{\cal B}^\prime}(x_2) J_\mu(x_1)\overline \chi^{\alpha}_{\cal B}(0))|\Omega \rangle,
\end{equation}
with ${\cal B},{\cal B}^{\prime}\equiv \Xi_c$ or $\Xi_c^\prime$. Here $t_1$ is the time when the electromagnetic current is inserted, $t_2$ is the time when the final baryon is annihilated and $\Gamma_4 = \frac{1}{2}
  \begin{bmatrix}
    1 & 0  \\
    0 & 0 
  \end{bmatrix}$ with $\Gamma_i = \frac{1}{2} \begin{bmatrix}
    \sigma_i & 0  \\
    0 & 0 
  \end{bmatrix}$.

The baryon interpolating fields are chosen as
  \begin{align}
  \chi^{}_{\Xi_c'}&= \frac{1}{\sqrt{2}} \varepsilon_{abc} \big[ \big( \ell^{T}_a (C \gamma _5) c_b \big)s_c + \big( s^{T}_a (C \gamma _5) c_b \big) \ell_c \big],\\
  \chi^{}_{\Xi_c} &= \frac{1}{\sqrt{6}} \varepsilon_{abc} \big[2 \big( s^{T}_a (C \gamma _5) \ell_b \big)c_c + \big( s^{T}_a (C \gamma _5) c_b \big) \ell_c - \big( \ell^{T}_a (C \gamma _5) c_b \big) s_c \big],
  \end{align}
where $\ell=u$ for the charged states $\Xi_c^+$, $\Xi_c^{'+}$, and $\ell= d$ for the neutral states $\Xi_c^0$, $\Xi_c^{'0}$. The indices $a,b,c$ denote color and the charge conjugation matrix is defined as $C= \gamma_4 \gamma_2$.

In the broken flavor SU(3) symmetry, there is a mixing between $\Xi_c$ and $\Xi_c^\prime$ baryons. Such mixing has been argued to be negligibly small~\cite{Franklin:1996ve, Aliev:2010ra}, which was also confirmed by lattice simulations~\cite{Brown:2014ena}. The reason for $\Xi_c$-$\Xi_c^\prime$ mixing being so small is the approximate SU(3) flavor symmetry and the heavy-quark spin symmetry, where the quantum numbers of the light degrees of freedom are exactly conserved. Therefore in our calculations we neglect the small mixing effects.

We use the following ratio to eliminate the normalization factors and to extract the baryon electromagnetic form factors
\begin{align}
	\begin{split}
    R(t_2,t_1,p',p,\Gamma,\mu) = &\frac{\langle F^{\cal B' J_\mu B}_{}(t_2,t_1;p',p;\Gamma) \rangle }{\langle F^{\cal B'B'}_\text{shwl}(t_2;p';\Gamma_4) \rangle }\\ &\times\Bigg[\frac{\langle F^{\cal BB}_\text{shsh}(t_2-t_1;p;\Gamma_4) \rangle \langle F^{\cal B'B'}_\text{shsh}(t_1;p';\Gamma_4) \rangle \langle F^{\cal B'B'}_\text{shsh}(t_2;p';\Gamma_4) \rangle }{\langle F^{\cal B'B'}_\text{shsh}(t_2-t_1;p';\Gamma_4) \rangle \langle F^{\cal BB}_\text{shsh}(t_1;p;\Gamma_4) \rangle \langle F^{\cal BB}_\text{shsh}(t_2;p;\Gamma_4) \rangle }\Bigg]^{\frac{1}{2}}.
\label{eq:Ratiospinhalf}
	\end{split}
\end{align}
In the large time limit, $t_2-t_1\gg a$ and $t_1\gg a$, the time dependence of the correlators are eliminated and the ratio in Eq.~\eqref{eq:Ratiospinhalf} reduces to
  \begin{equation}
     R(t_2,t_1;p',p;\Gamma;\mu)  \xrightarrow[t_1\gg a]{t_2 -t_1 \gg a} \Pi(p',p;\Gamma;\mu).
     \label{eq:RtoPi}
  \end{equation}
The Sachs form factors can be extracted from the final form of the ratio above by choosing specific combinations of the projection matrices $\Gamma$ and the Lorentz index $\mu$:
  \begin{equation}
      \Pi(p',p;\Gamma_4;\mu=4) = \frac{1}{2}\sqrt{\frac{(E_{{\cal B}'}+m_{{\cal B}'})(E_{\cal B}+m_{\cal B})}{E_{\cal B} E_{{\cal B}'}}} G_E(q^2), 
     \label{eq:gEtrans}
  \end{equation}
  \begin{equation}
     \Pi(p',p;\Gamma_j;\mu=i) =  \frac{\epsilon_{ijk} q_k}{2} \sqrt{\frac{(E_{{\cal B}}+m_{{\cal B}})}{E_{\cal B}E_{{\cal B}'}(E_{{\cal B}'}+m_{{\cal B}'})}} G_M(q^2). 
     \label{eq:gMtrans}
  \end{equation}
Note that when the incoming and the outgoing baryon states are identical, \emph{i.e.} ${\cal B}^\prime={\cal B}$, the electric form factor $G_E (q^2=0)$ gives the electric charge of the baryon. As for the magnetic form factor, $G_M(q^2=0)$ gives the magnetic moment of the baryon when ${\cal B}^\prime={\cal B}$ and it gives the transition magnetic moment when ${\cal B}^\prime \neq {\cal B}$.
\subsection{Lattice Setup}
We use gauge configurations generated by the PACS-CS collaboration~\cite{Aoki:2008sm}, with $O(a)$-improved Wilson quark action and the Iwasaki gauge action. We run our simulations with the hopping parameter of the light quarks $\kappa^\ell = 0.13781$ which gives a near physical value of pion mass $ m_\pi \approx 156$~MeV. We use the Clover action also for the strange valence quark and take its hopping parameters to be equal to that of the strange sea quark, $\kappa^s_\text{val}=\kappa^s_\text{sea} = 0.13640$. Further details of the gauge configurations are given in Table~\ref{table:gauge}. 
\begin{table}[!htb]
\caption{The details of the gauge configurations we employ~\cite{Aoki:2008sm}. $N_s$ and $N_t$ are the spatial and temporal sizes of the lattice, respectively, $N_f$ is the number of flavors, $a$ is the lattice spacing, $L$ is the volume of the lattice, $\beta$ is the inverse gauge coupling, $c_{sw}$ is the Clover coefficient, $\kappa_{sea}^f$ is the hopping parameter of the quark with flavor $f$ and $m_\pi$ is the pion mass.}
\centering
	\setlength{\extrarowheight}{7pt}
\begin{tabular*}{0.8\textwidth}{@{\extracolsep{\fill}}|l|l|l|l|l|l|l|l|l|l|}
     \hline
 $N_s \times N_t$ & $N_f$  &  $a$ (fm) & $L$ (fm) & $\beta$ & $c_{sw}$& $\kappa_{sea}^\ell$  & $\kappa_{sea}^s$ & $\#$ of conf. & $m_\pi$[MeV]\\
 \hline
 $32^3 \times 64$ & 2+1 & 0.0907(13)  &  2.90 & 1.90 & 1.715 & 0.13781 & 0.13640 & 163 & 156(7)(2)\\
  \hline
\end{tabular*}
\label{table:gauge}
\end{table}

As for the charm quarks, we apply a Clover action in the form used by Fermilab \cite{ElKhadra:1996mp}, MILC Collaborations \cite{Burch:2009az, Bernard:2010fr} and in our previous work~\cite{Bahtiyar:2015sga}. In this action, the Clover coefficients in the action are set to tadpole-improved value $\frac{1}{u_0^3}$ where $u_0$ is the average link. We follow the approach in \cite{Mohler:2011ke} and estimate $u_0$ to be the fourth root of the average plaquette. We use the value of the hopping parameter $\kappa_c= 0.1246$ as determined in our previous work by tuning the spin-averaged static masses of charmonium and open charm mesons to their experimental values~\cite{Can:2013tna}. 

In order to increase statistics, we insert positive and negative momenta in all spatial directions and make a simultaneous fit over all data. We also take account of current insertion along all spatial directions. The source-sink time separation is fixed to 12 lattice units (1.09 fm), which is enough to avoid excited state contaminations for electromagnetic form factors \cite{Can:2013tna}. We have employed multiple source-sink pairs by shifting them 12 lattice units in the temporal direction. All statistical errors are estimated by the single-elimination jackknife analysis. We insert momentum up to nine units: $(|p_x|, |p_y|, |p_z|)=$(0,0,0), (1,0,0), (1,1,0), (1,1,1), (2,0,0), (2,1,0), (2,1,1), (2,2,0), (2,2,1) and average over equivalent momenta. We consider point-split lattice vector current
  \begin{equation}
		 j_\mu = \frac{1}{2} [\overline{q}(x+\mu) U_\mu^{\dagger}(1+\gamma_\mu)q(x)-\overline{q}(x)U_\mu(1-\gamma_\mu)q(x+\mu)],
		 \label{eq:pointsplit}
  \end{equation}
which is conserved by Wilson fermions therefore do not need any renormalization.
  
We use wall-source/sink method \cite{Can:2012tx} that provides a simultaneous extraction of all spin, momentum and projection components of the correlators. On the other hand the wall source/sink is a gauge-dependent object which requires fixing the gauge. We fix the gauge to Coulomb, which gives a somewhat better coupling to the ground state than Landau. By using the wall method we can first compute the shell and wall propagators regardless of the current and momenta inserted. Then we contract the propagators to obtain the three-point correlators. Only the connected diagrams are considered in this work. Possible effects of the disconnected diagrams are discussed in the following section. 

We performed our computations using a modified version of Chroma software system~\cite{Edwards:2004sx} on CPU clusters and with QUDA~\cite{Babich:2011np,Clark:2009wm} for propagator inversion on GPUs.

\section{Results And Discussion}
\label{sec:results}
First, we discuss our results for the $\Xi_c$ and $\Xi_c'$ masses. We extract the ground-state masses using the two-point correlation functions in Eq.~\eqref{eq:twopt}. Our results are given in Table~\ref{tab:mass}, along with their experimental values and those of other lattice collaborations. As our results are obtained with near physical values of light-quark masses, we do not make any chiral extrapolations. Our results for the baryon masses differ by only ~$2\%$ as compared to the experimental values, while there is a good agreement for the mass splitting $m_{\Xi_c'} - m_{\Xi_c}$. Note that it is tempting to attribute this small discrepancy to Clover action we are employing for the charm quarks, however, it has been confirmed in Ref.~\cite{Can:2015exa} that the mass of the triply charmed $\Omega_{ccc}$ baryon can be calculated in very good agreement with other lattice determinations using relativistic heavy-quark actions. Such small discrepancy may be due to our choosing $\kappa^s_\text{val}=\kappa^s_\text{sea} = 0.13640$ to be consistent with PACS-CS. This choice of $\kappa$ values leads to an overestimation of the $\Omega\,{\rm (sss)}$ mass around $100$ MeV as compared to its experimental value~\cite{Aoki:2008sm} and of the Kaon mass~\cite{Menadue:2011pd}. On the other hand, the form factor determinations are rather insensitive to mild changes in baryon masses at the current precision level and a discrepancy of $2\%$ can be safely neglected. 

\begin{table}[t]
\centering
\caption{The $\Xi_c$ and $\Xi_c'$ masses together with experimental values and those of other lattice collaborations.}
\label{tab:mass}
	\setlength{\extrarowheight}{7pt}
\begin{tabular*}{0.8\textwidth}{@{\extracolsep{\fill}}|c|c|c|c|c|c|}
\hline
	 & This work & PACS-CS \cite{Namekawa:2013vu} & ETMC \cite{Alexandrou:2014sha} & Briceno et al. \cite{Briceno:2012wt} & Experiment \cite{Olive:2016xmw}\\ \hline \hline
 $m_{\Xi_c}$ [GeV]& 2.519(15) &2.455(16)&2.469(28)&2.439(29)(25)(7)&2.470 (1)\\ \hline
 $m_{\Xi_c'}$ [GeV]& 2.646(17) &2.583(20)&2.542(27)&2.568(25)(12)(6)&2.577 (3)\\ \hline
 \end{tabular*}
\end{table}

In this work we focus on the magnetic form factors and we make an extrapolation to zero momentum transfer in order to obtain the magnetic moments. While the electric charge $G_E(0)$ can be computed directly with our formulation on the lattice, we cannot make a direct measurement of the magnetic form factor at zero momentum $G_M(0)$. To this end, we use the following dipole form to describe the $Q^2$ dependence of the form factors:
\begin{equation}
G_{E,M} (Q^2) = \dfrac{G_{E,M}(0)}{\left(1+Q^2/\Lambda_{E,M}^2\right)^2}.
\label{eq:dipoleform}
\end{equation}
The form factors, can be calculated from individual quark contributions by
\begin{equation}
		G_{M} (Q^2) = \frac{2}{3} G_{M}^c (Q^2) - \frac{1}{3} G_{M}^s (Q^2) + c_\ell G_{M}^\ell (Q^2),
		\label{eq:individual}
\end{equation}
where $c_\ell=-1/3$ for the $d$ quark and $c_\ell=2/3$ for the $u$ quark. We combine the individual quark contributions using Eq.~\eqref{eq:individual} for each momentum transfer $Q^2$ and extrapolate the combined form factor values to $Q^2=0$. 

\begin{table}[ht]
\caption{Individual contributions of $u$/$d$, $s$ and $c$ quarks to the magnetic form factor at different $Q^2$ values of all transitions we study.}
\label{tab:quarkindall}
\setlength{\extrarowheight}{5pt}
\begin{tabular}{|c|c||c|c|c|c|c|c|c|c|}
\hline
 
 \multicolumn{2}{|c||}{$Q^2$ [GeV$^2$]} & 0 & 0.181 & 0.360 & 0.536 & 0.710 & 0.882 & 1.052 & 1.385\\ \hline \hline
 \multirow{3}{*}{$\Xi_c \gamma \rightarrow \Xi_c $} & $u$ contr. & 0.209(272) & 0.189(143) & 0.038(67) & 0.143(69) & -0.053(66) & -0.028(45) & 0.023(32) & 0.032(34)\\ \cline{2-10}
  &$s$ contr.& 0.195(91) & 0.156(65) & 0.110(50) & 0.096(45) & 0.120(41) & 0.055(33) & 0.052(31) & 0.053(27)\\ \cline{2-10}
  &$c$ contr.& 0.912(39) & 0.885(37) & 0.853(37) & 0.824(39) & 0.796(44) & 0.774(45) & 0.752(49) & 0.723(59)\\\hline \hline
 \multicolumn{2}{|c||}{$Q^2$ [GeV$^2$]} & 0 & 0.181 & 0.360 & 0.537 & 0.712 & 0.885 & 1.056 & 1.391\\ \hline \hline
 \multirow{3}{*}{$\Xi'_c \gamma \rightarrow \Xi'_c $} & $u$ contr. & 2.514(500) & 1.703(421) & 2.11(341) & 1.515(245) & 1.081(229) & 1.098(162) & 0.921(144) & 0.772(153)\\ \cline{2-10}
 &$s$ contr.& 2.214(253) &1.913(200) & 1.559(162) & 1.348(148) & 1.233(146) & 1.054(127) & 0.923(116) & 0.804(108)\\ \cline{2-10}
 &$c$ contr.& -0.249(47) &-0.241(468) & -0.241(46) & -0.242(47) & -0.213(50) & -0.224(48) & -0.232(49) & -0.209(52)\\
  \hline \hline
 \multicolumn{2}{|c||}{$Q^2$ [GeV$^2$]} & 0 & 0.174 & 0.362 & 0.547 & 0.730 & 0.911 & 1.089 & 1.439\\ \hline \hline
 \multirow{3}{*}{$\Xi_c \gamma \rightarrow \Xi'_c $} & $u$ contr. & 2.057(359) & 1.696(304) & 1.392(209) & 1.040(176) & 1.062(171) & 0.921(133) & 0.783(105) & 0.528(103)\\ \cline{2-10}
 &$s$ contr.& -1.943(223) & -1.669(179) & -1.378(148) & -1.169(127) & -0.998(109) & -0.883(99) & -0.783(93) & -0.670(85)\\ \cline{2-10}
 &$c$ contr.& 0.023(30) & 0.022(29) & 0.021(29) & 0.018(29) & 0.031(30) & 0.026(30) & 0.019(31) & 0.027(32)\\ \hline
\end{tabular}
\end{table}

\begin{figure}[t]
    \caption{$Q^2$ dependence of the magnetic form factors of $\Xi_c^+ \gamma \rightarrow \Xi_c^+ $, $\Xi_c'^+ \gamma \rightarrow \Xi_c'^+$ and $\Xi_c^+ \gamma \rightarrow \Xi_c'^+$}
    \centering
      \includegraphics[width=0.99\textwidth]{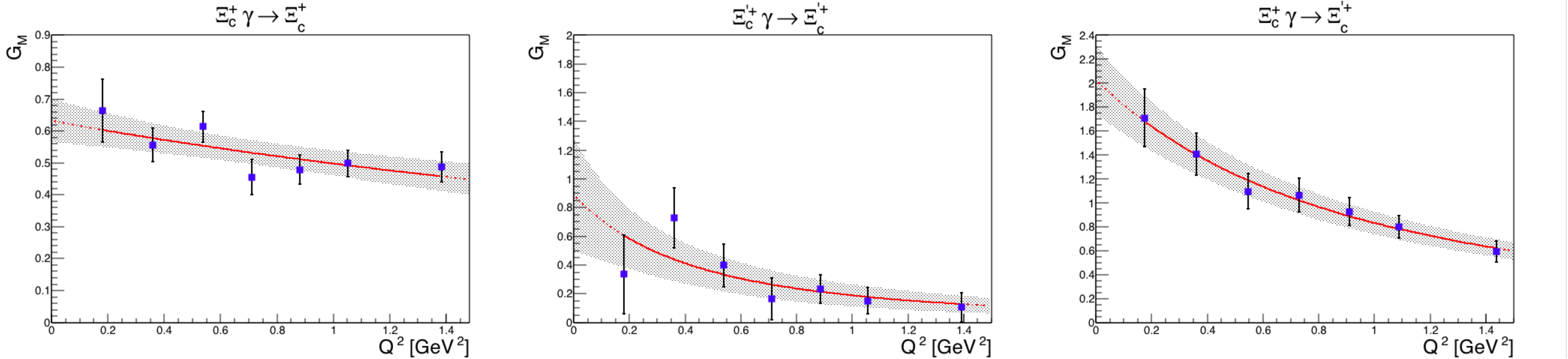}
	  \label{fig:gmzerousc}
  \end{figure}

\begin{figure}[t]
    \caption{$Q^2$ dependence of the magnetic form factors of $\Xi_c^0 \gamma \rightarrow \Xi_c^0 $, $\Xi_c'^0 \gamma \rightarrow \Xi_c'^0$ and $\Xi_c^0 \gamma \rightarrow \Xi_c'^0$}
    \centering
      \includegraphics[width=0.99\textwidth]{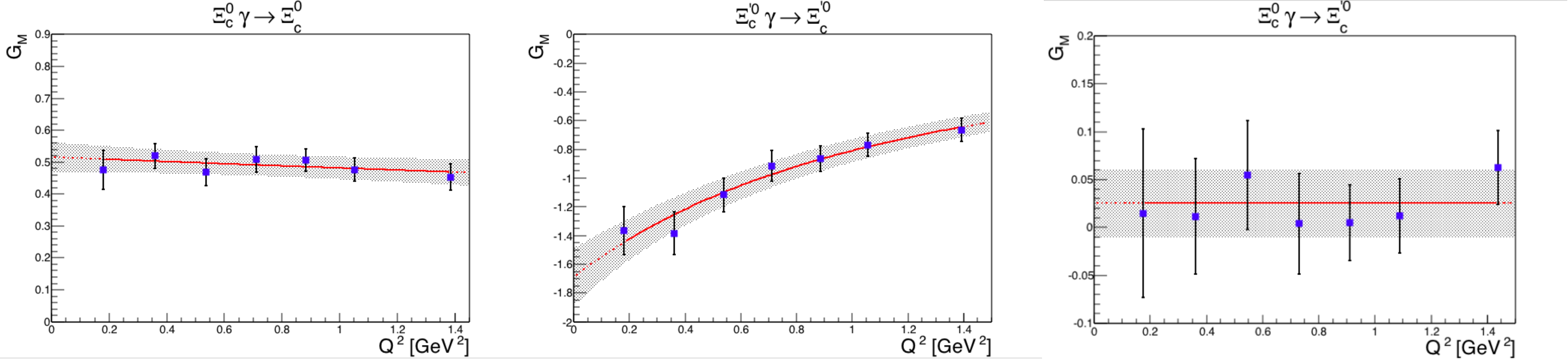}
	  \label{fig:gmzerodsc}
  \end{figure}

In the left two panels of Fig.~\ref{fig:gmzerousc} and Fig.~\ref{fig:gmzerodsc}, we plot the magnetic form factors $G_M(Q^2)$ of the charged $\Xi_c^+$, $\Xi_c^{\prime +}$ and neutral $\Xi_c^0$, $\Xi_c^{\prime 0}$ states as functions of $Q^2$. In the SU(4) limit all individual valence quarks give equal contributions to $\Xi_c^\prime$ form factors, similar to proton. However, SU(4) symmetry is badly broken and the quark contribution decreases as the quark mass increases. This is consistent with what has been observed in previous works on charmed baryons~\cite{Can:2012tx, Can:2013tna, Can:2015exa, Bahtiyar:2015sga}. This is also evident in Table~\ref{tab:quarkindall}, where we list individual contributions of $u$/$d$, $s$ and $c$ separately for the transitions we study at all $Q^2$. The heavy $c$-quark contribution is one order smaller than those of light $u$/$d$ and $s$ quarks. This dominance of light quarks yields a soft core and the form factor decreases rapidly as $Q^2$ increases. Due to flavor asymmetry of $\Xi_c$ baryon wavefunctions, the $u/d$- and $s$-quark contributions cancel each other to a great extent leading to a form factor that is dominantly determined by the $c$ quark. This cancellation would be exact in the SU(3) symmetric limit. Since the $Q^2$ dependence of the $\Xi_c$ form factors is controlled by the heavy $c$ quark, it yields a hard core and the form factor decreases less rapidly as $Q^2$ increases. 

The light $u/d$- and $s$-quark contributions to the transition magnetic form factors of $\Xi_c \gamma \rightarrow \Xi_c^{'}$ are equal in magnitude and opposite in sign. On the other hand, the $c$ quark has almost no effect. When the quark contributions are combined using the formula in Eq.~\eqref{eq:individual}, the $u$ and $s$ contributions to $\Xi_c^+ \gamma \rightarrow \Xi_c^{'+}$ are multiplied with electric charges of opposite sign and add constructively. In contrast, the neutral transition $\Xi_c^0 \gamma \rightarrow \Xi_c^{'0}$ is highly suppressed as a result of equal electric charges of the $d$ and $s$ quarks. According to conserved U-spin flavor symmetry, which assumes a degeneracy between two equally charged $d$ and $s$ quarks, a transition from $\Xi_c^0$ to $\Xi_c^{'0}$ is forbidden. Our results are in agreement with what U-spin flavor symmetry predicts. As shown in Fig.~\ref{fig:gmzerodsc} the  magnetic form factor of $\Xi_c^0 \gamma \rightarrow \Xi_c^{'0}$ neutral transition is consistent with zero. 

In this work, we neglect the effects of disconnected diagrams, which are noisy and costly to compute. Contributions of disconnected diagrams to isovector electromagnetic form factors are usually suppressed. We expect the sea-quark effects to be also suppressed in our results. In the case of nucleon electromagnetic form factors, the contributions of disconnected diagrams have been found to be approximately $0.5\%$ of those of connected diagrams~\cite{Constantinou:2014tga}. They may however play an important role for the $\Xi_c^0 \gamma \rightarrow \Xi_c^{'0}$ transition, where the connected contributions cancel.

As a rule of thumb, the finite-size effects should be negligible when $m_\pi L \geq 4$. The $\kappa^\ell = 0.13781$ configurations that we employ, on the other hand, yield $m_\pi L = 2.3$ which is below the empirical bound. We have, however, confirmed that the finite-size effects on this particular setup are under control for physical quantities related to strange and charmed baryons~\cite{Can:2015exa}. As we have discussed above, the magnetic form factor of $\Xi_c^0 \gamma \rightarrow \Xi_c^{'0}$ should vanish due to U-spin flavor symmetry which assumes a degeneracy between $d$- and $s$-quarks. This is realized in our numerical calculations when the $d$- and $s$-quark contributions cancel each other so that the magnetic form factor is consistent with zero as shown in Fig.~\ref{fig:gmzerodsc}. This indicates that the the finite-size effects on the light quarks are either similar as compared to those of strange and charmed quarks or any unaccounted effect is already hidden in the statistical error of the quantities that we extract. Note that, our discussion here gives only a qualitative account of the possible finite-size effects rather than a quantitative estimation, which requires further investigation. 

\begin{table}[t]
\centering
\caption{The combined form factor as obtained using Eq.~\eqref{eq:individual} and extrapolated to $Q^2=0$, together with the magnetic moments in units of nuclear magneton.}
\label{tab:indquark}
	\setlength{\extrarowheight}{7pt}
\begin{tabular}{|c||c|c|}
\hline
Transition  & $G_{M} (0)$ & Magnetic moment[$\mu_N$]\\ \hline \hline
 $\Xi_c^+ \gamma \rightarrow \Xi_c^+ $ &   $0.631(68)$ & $0.235(25)$\\ \hline
 $\Xi_c^0 \gamma \rightarrow \Xi_c^0 $ & $0.516(46)$ & $0.192(17)$\\ \hline
 $\Xi_c^{'+} \gamma \rightarrow \Xi_c^{'+}$ & $0.889(397)$ & $0.315(141)$\\ \hline
 $\Xi_c^{'0} \gamma \rightarrow \Xi_c^{'0}$&  $-1.689(201)$ & $-0.599(71)$\\ \hline
 $\Xi_c^+ \gamma \rightarrow \Xi_c^{'+}$&  $2.027(286)$ & $0.729(103)$\\ \hline
 $\Xi_c^0 \gamma \rightarrow \Xi_c^{'0}$  & $0.025(36)$ & $0.009(13)$\\ \hline
\end{tabular}
\end{table}
  
Using the values of the magnetic form factors at $Q^2=0$ in Table~\ref{tab:indquark}, we calculate the magnetic moments in nuclear magnetons by using
 \begin{equation}
	\mu_B = G_M(0) (e/2 m_B) = G_M(0) (m_N/m_B) \mu_N,
	\label{eq:nuclearmagneton}
 \end{equation}
where $m_N$ is the physical nucleon mass and $m_B$ is the baryon mass obtained on the lattice. In Table~\ref{tab:indquark}, we list the combined form factor as obtained using Eq.~\eqref{eq:individual} and extrapolated to $Q^2=0$, that is $G_{M} (0)$, the magnetic moments calculated using $G_{M} (0)$ in units of nuclear magneton.

The decay width of $\Xi_c^\prime$ baryon is related to the Pauli form factor $F_2(0)$ of $\Xi_c \gamma \rightarrow \Xi_c^{'}$: 
\begin{equation}
		 \Gamma_{B \gamma \rightarrow B' } = \frac{4 \alpha |\vec{q}|^3}{(m_{B'}+m_B)^2}|F_2(0)|^2 \quad \text{with}\quad |\vec{q}| = \frac{(m_{B'}^2-m_B^2)}{2 m_{B'}}.
		 \label{eq:decayrate}
\end{equation}
Since the relation in Eq.~\eqref{eq:decayrate} is defined in the continuum, we evaluate it by using the experimental masses of $\Xi_c$ and $\Xi_c^{'}$. In order to extract $F_2(0)$ from the Sachs form factors $G_E(Q^2)$ and $G_M(Q^2)$, we solve the two equations in Eqs.~\eqref{elS} and~\eqref{magS} simultaneously for all lattice data and extrapolate to $Q^2=0$. At zero momentum transfer, $G_E(0)=F_1(0)$ and we can immediately deduce that $F_1(0)$ must have a very small value, if not zero. Since $\Xi_c \gamma \rightarrow \Xi_c^{'}$ cannot occur through electric transition, this implies $G_M(0)\simeq F_2(0)$. Consistently, we find
\begin{align}
	\begin{split}
	F_2(0)=2.036(280)\quad \text{for} &\quad \Xi_c^+ \gamma \rightarrow \Xi_c'^+, \\
	F_2(0)=0.039(46)\quad \text{for} &\quad \Xi_c^0 \gamma \rightarrow \Xi_c'^0.
\end{split}
\end{align}
Using the formula in Eq.~\eqref{eq:decayrate}, we obtain the decay widths of $\Xi_c$ baryons as follows:
\begin{equation}
	\Gamma_{\Xi_c'^+}=5.468(1.500)~\text{keV}, \quad \Gamma_{\Xi_c'^0}=0.002(4)~\text{keV}.
\end{equation} 
The decay width can be translated into a lifetime using $\tau = \frac{1}{\Gamma}$;
\begin{equation}
	\tau_{\Xi_c'^+} = 1.148(322)\times10^{-19}~\text{s}.
\end{equation}

Both neutral and charged transitions of $\Xi_c \gamma \rightarrow \Xi'_c $ have been previously studied using QCD sum rules \cite{Aliev:2016xvq}, heavy hadron chiral perturbation theory \cite{Cheng:1992xi,Banuls:1999br,Jiang:2015xqa}, quark model \cite{Ivanov:1999bk,Dey:1994qi} and bag model \cite{Bernotas:2013eia}. For the charged transition, our lattice results for the transition form factor and decay width are in agreement with those from QCD sum rules~\cite{Aliev:2016xvq}, while other methods previously used predict higher values. In the case of neutral transition, their predictions for the transition form factors are small but finite, while we find no signal on the lattice.

\section{Summary and Conclusion}
We studied the magnetic form factors of the $\Xi_c \gamma \rightarrow\Xi^\prime_c$ transition and of the $\Xi_c$, $\Xi_c^\prime$ baryons in 2+1-flavor lattice QCD. We have extracted the magnetic Sachs and Pauli form factors which give the $\Xi_c$-$\Xi_c^\prime$ transition magnetic moment and the decay widths of $\Xi_c^\prime$ baryons. We determined individual quark contributions to the magnetic moments, which give an invaluable insight to the dynamics of $u/d$, $s$ and $c$ quarks having masses at different scales. In the case of $\Xi'_c$ baryons the heavy $c$-quark contribution is much smaller than those of light $u/d$ and $s$ quarks. On the other hand, due to antisymmetric flavor wavefunctions of the $\Xi_c$ baryons, the $u/d$- and $s$-quark contributions cancel each other to a great extent leading to a form factor that is dominantly determined by the $c$ quark. We find that the $c$ quark has a negligibly small contribution to the $\Xi_c \gamma \rightarrow \Xi^\prime_c$ transition and, $u/d$ and $s$ quarks contribute with opposite sign. Using the Pauli form factor $F_2(Q^2=0)$, we extracted the decay widths of $\Xi_c^\prime$ baryons. The decay width of the charged $\Xi_c^\prime$ baryon on the lattice is determined as $\Gamma_{\Xi_c'^+}=5.468(1.500)~\text{keV}$ and we did not find a signal for the magnetic form factor of the neutral transition $\Xi_c^0 \gamma \rightarrow\Xi_c^{\prime 0}$, which is suppressed by the U-spin flavor symmetry.

\acknowledgments
This work is supported in part by The Scientific and Technological Research Council of Turkey (TUBITAK) under project number 114F261 and in part by KAKENHI under Contract Nos. 25247036, 24250294 and 16K05365. This work is also supported by the Research Abroad and Invitational Program for the Promotion of International Joint Research, Category (C) and the International Physics Leadership Program at Tokyo Tech.

%


\end{document}